\documentclass{aa}  
\usepackage{color}
\usepackage{graphicx}
\usepackage{lscape}
\usepackage{natbib}
\usepackage[varg]{txfonts}
\usepackage{url}
\usepackage{xspace}
\bibpunct{(}{)}{;}{a}{}{,}
\def\etal{{et\,al.}\ }

\newcommand{\Teff}{$T\mathrm{\hspace*{-0.4ex}_{eff}}$\,}
\newcommand{\logg}{$\log\,g$\hspace*{0.5ex}}

\def\elf{PG\,1159$-$035}
\def\pgvier{PG\,1424+535}
\def\pgfuenf{PG\,1520+525}
\def\pgev{PG\,1144+005}
\def\pgsieben{PG\,1707+427}

\begin{document}

\title{The far-ultraviolet spectra of two hot PG\,1159 stars}

\author{
K\@. Werner\inst{1} \and 
        T\@. Rauch\inst{1} \and
        J\@. W\@. Kruk\inst{2} 
}

\institute{Institute for Astronomy and Astrophysics, Kepler Center for Astro and
Particle Physics,  Eberhard Karls University, Sand~1, 72076
T\"ubingen, Germany\\ \email{werner@astro.uni-tuebingen.de}\and
           NASA Goddard Space Flight Center, Greenbelt, MD\,20771, USA}

\date{Received xx xx 2016 / Accepted xx xx 2016}

\authorrunning{K. Werner \etal}
\titlerunning{The far-ultraviolet spectra of two hot PG\,1159 stars}

\abstract{PG\,1159 stars are hot, hydrogen-deficient (pre-) white
  dwarfs with atmospheres mainly composed of helium, carbon, and
  oxygen. The unusual surface chemistry is the result of a late
  helium-shell flash. Observed element abundances enable us to test
  stellar evolution models quantitatively with respect to their
  nucleosynthesis products formed near the helium-burning shell of the
  progenitor asymptotic giant branch stars. Because of the high
  effective temperatures ($T\mathrm{\hspace*{-0.4ex}_{eff}}$),
  abundance determinations require ultraviolet spectroscopy and
  non-local thermodynamic equilibrium model atmosphere analyses. Up to
  now, we have presented results for the prototype of this spectral class
  and two cooler members (\Teff\ in the range
  85\,000--140\,000~K). Here we report on the results for two even
  hotter stars  (\pgfuenf\ and \pgev, both with \Teff = 150\,000~K)
  which are the only two objects in this temperature--gravity region for
    which useful far-ultraviolet spectra are available, and revisit
  the prototype star. Previous results on the abundances of some
    species are confirmed, while results on others (Si, P, S) are
    revised. In particular, a solar abundance of sulphur is measured in
    contrast to earlier claims of a strong S deficiency that
    contradicted stellar evolution models. For the first time, we
    assess the abundances of Na, Al, and Cl with newly constructed
    non-LTE model atoms. Besides the main constituents (He, C, O), we
    determine the abundances (or upper limits) of N, F, Ne, Na,
    Al, Si, P, S, Cl, Ar, and Fe. Generally, good agreement with
  stellar models is found.}

\keywords{
          stars: abundances -- 
          stars: atmospheres -- 
          stars: evolution  -- 
          stars: AGB and post-AGB --
          white dwarfs}

\maketitle
%

\begin{figure*}[bth]
 \centering  \includegraphics[width=1.0\textwidth]{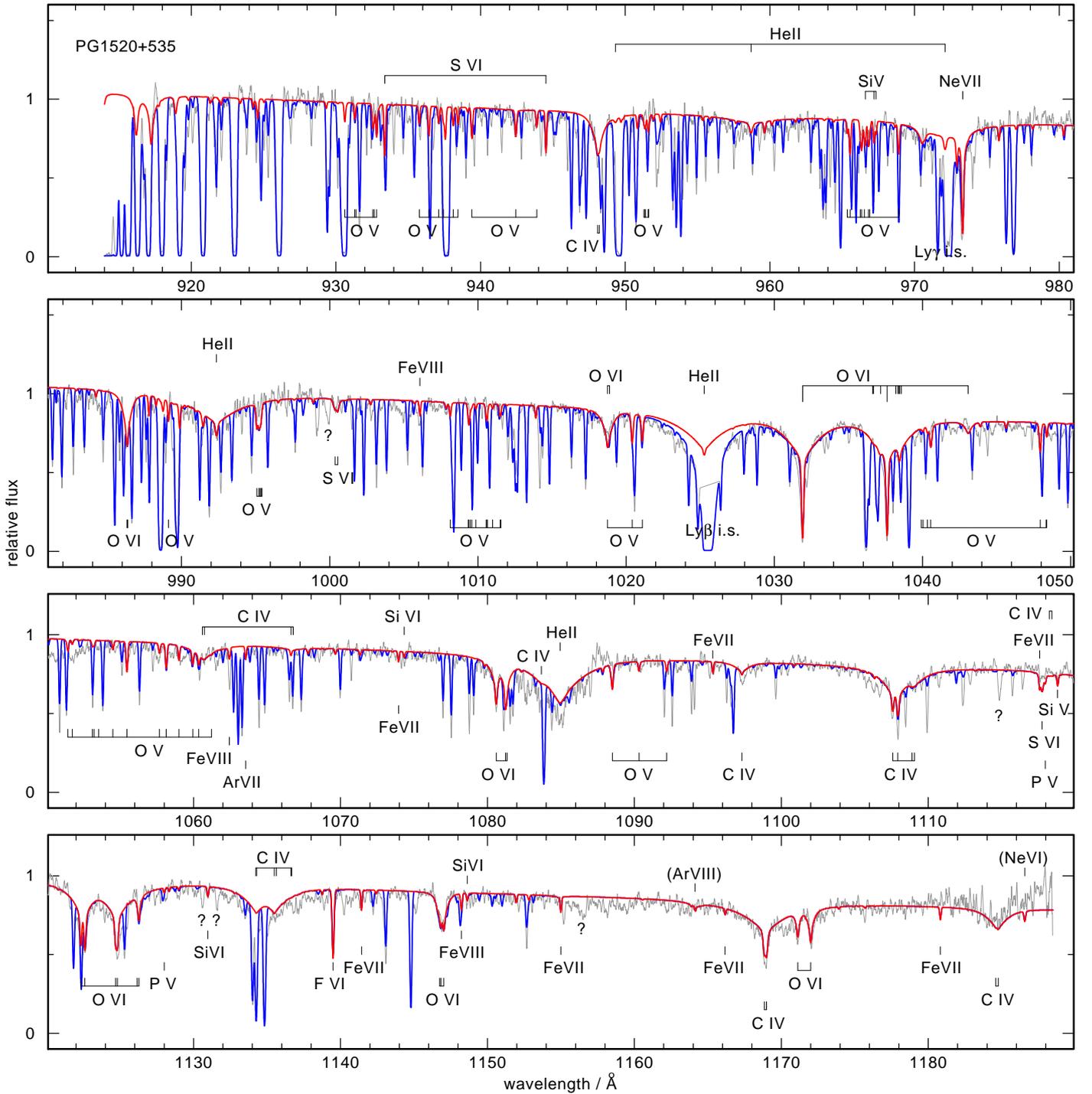}
  \caption{FUSE spectrum of \pgfuenf\ (thin black line) compared to a
    photospheric model spectrum (thick red line) with the finally
    adopted parameters as listed in Table\,\ref{tab:stars}. Prominent
    photospheric spectral lines are identified. Uncertain
    identifications are given in brackets. Unidentified, probably
    photospheric, lines are indicated by question marks. Geocoronal
    emission in the center of Lyman~$\beta$ was cut
    out. Overplotted in blue is the same model including ISM lines.}\label{fig:pg1520_fuse}
\end{figure*}

\begin{figure*}[bth]
 \centering  \includegraphics[width=1.0\textwidth]{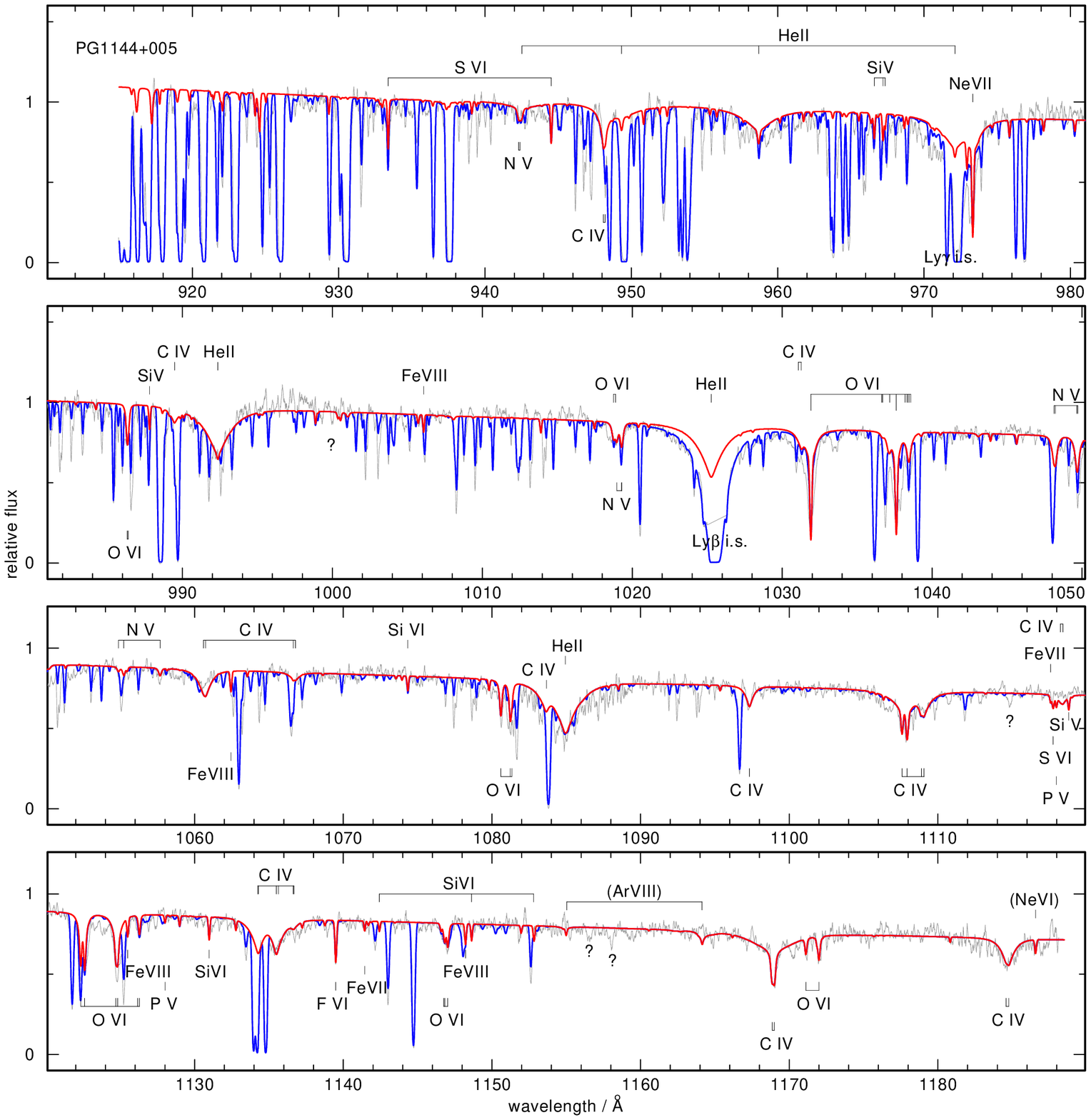}
  \caption{As Fig.\,\ref{fig:pg1520_fuse}, for \pgev.}\label{fig:pg1144_fuse}
\end{figure*}

\section{Introduction}
\label{intro}

PG\,1159 stars are hydrogen-deficient and occupy the hot end of the
white dwarf cooling sequence \citep[effective temperatures ranging
  between \Teff = 75\,000~K and
  250\,000~K,][]{2006PASP..118..183W,2014A&A...564A..53W,2015A&A...584A..19W}.
Their surfaces are dominated by helium-rich intershell matter dredged
up by a late helium-shell flash (thermal pulse, TP). The main
atmospheric constituents are helium and carbon (He = 0.30--0.92, C =
0.08--0.60, mass fractions), often accompanied by large amounts of
oxygen (up to 0.20). The high abundances of C and O are witness to
efficient overshoot of intershell convection in asymptotic giant
branch (AGB) stars, drawing these elements from the C--O stellar core
\citep{1999A&A...349L...5H,2000A&A...360..952H,2016A&A...588A..25M}.
Measured overabundances of neon confirm this picture. Our
investigation of these stars  now concentrates on the trace
elements up to the iron group. Quantitative spectroscopic abundance
determinations can be compared to predictions from nucleosynthesis
calculations in stellar evolution models for AGB stars.

The first comprehensive analysis of trace metal abundances was
performed for the prototype of the PG\,1159 spectroscopic class,
\elf\ \citep[\Teff = 140\,000~K,][]{2007A&A...462..281J}. Then, two
cooler objects from this group were analysed \citep[\pgvier\ and
  \pgsieben, \Teff = 110\,000 and
  85\,000~K,][]{2015A&A...582A..94W}. 

Prerequisites for these types of work are ultraviolet (UV) spectroscopic
observations and model atmospheres accounting for departures from
local thermodynamic equilibrium (LTE), because of the high effective
temperatures. The far-UV spectral range covered by the Far Ultraviolet
Spectroscopic Explorer (FUSE, 912--1188~\AA) is a most rewarding source.

In the present paper, we assess the element abundances of two
representatives of this group that are hotter than the prototype,
namely 150\,000~K: \pgfuenf\ and \pgev. They are the hottest PG\,1159
stars for which useful FUSE data are available. For their analysis we
developed new non-LTE model atoms, concentrating on high ionisation
stages, particularly for elements not considered previously (Na, Mg,
Al, Cl). We strive to identify spectral lines that have not been
observed before in any stellar spectrum. The effective temperature of
both stars is only rivaled by a few low-gravity PG\,1159-type central
stars with \Teff\ up to 170\,000\,K \citep[K1--16,
    RX\,J2117.1+3412, NGC\,246, Longmore~4;][]{2010ApJ...719L..32W}
for which good FUSE spectra are available. But for a comprehensive
quantitative analysis, they require expanding-atmosphere models and the
design of model atoms for even higher ionisation stages. In addition
to the two program stars, we revisit the prototype itself. It is
necessary and useful to analyse a larger number of these stars
because, according to stellar models, the abundances of individual
trace elements strongly depend on the stellar mass.

\section{The program stars}
\label{sect:programstars}

Both stars analysed in this paper were discovered in the Palomar-Green
Survey \citep{1986ApJS...61..305G}. \pgfuenf\ was identified as a very
hot He-rich degenerate, classified a PG\,1159 star, and a parameter
estimate based on LTE models was attempted \citep[\Teff\,$\approx
  100\,000$\,K,  $\log$\,($g$/cm\,s$^{-2}$) $>
  6$,][]{1985ApJS...58..379W}. \pgev\ was classified a PG\,1159 star
by  \citet{1985BAAS...17..838B} and \citet{1986ApJS...61..305G}. Both
objects are non-pulsators \citep{1987ApJ...323..271G}, in contrast to
many other PG\,1159 stars, including the prototype (GW~Vir
variables). \pgfuenf\ is the central star of an old planetary nebula
\citep[PK\,085+52\,1,][]{1995AJ....110.1285J}.

\subsection{\pgfuenf}

The first non-LTE analysis was performed by
\citet{1991A&A...244..437W} using optical spectra. They found \Teff =
$140\,000\pm 14\,000$\,K, \logg = $7.0\pm 0.5$, He = 0.33, C = 0.50, O
= 0.17 (element abundances in mass fractions). Nitrogen lines were not
discovered and N $< 4\cdot10^{-3}$ was derived. Utilizing UV spectra
taken with the Goddard High Resolution Spectrograph aboard the
\emph{Hubble} Space Telescope, these parameters were improved by
\citet{1998A&A...334..618D}: \Teff = $150\,000\pm 7500$\,K, \logg =
$7.5\pm 0.5$, He = 0.44, C = 0.39, O = 0.17, N $<
1.5\cdot10^{-4}$. The slightly higher temperature was confirmed by the
analysis of a Chandra/LETG soft-X-ray spectrum \citep[$150\,000 \pm
  10\,000$\,K,][]{2012A&A...546A...1A}.

Abundances of further elements were measured from FUSE spectra. Ne =
0.02 and F = $10^{-4}$ was determined
\citep{2004A&A...427..685W,2005A&A...433..641W}. Preliminary results
for other light element abundances were presented by
\citet{2007ASPC..372..237R}: Si $< 3.6 \times 10^{-5}$, P = $6.4
\times 10^{-6}$, S = $5.0 \times 10^{-5}$. An upper abundance limit
for argon of roughly solar was derived \citep[Ar $< 8 \times
  10^{-5}$,][]{2007A&A...466..317W}. From the lack of \ion{Fe}{vii}
lines, an upper abundance limit of Fe $<$ 0.3 solar was estimated
\citep{2002A&A...389..953M}. Finally, from a fit to newly identified
\ion{Fe}{viii} lines, a solar iron abundance was claimed
\citep{2011A&A...531A.146W}. 

\subsection{\pgev}

The first non-LTE analysis was performed by
\citet{1991A&A...247..476W} using optical spectra. They found \Teff =
$150\,000\pm 15\,000$~K, \logg = $6.5\pm0.5$, He = 0.39, C = 0.58, O =
0.016. A relatively high nitrogen abundance was determined (N =
0.015). 

As in the case of \pgfuenf, abundances of further elements were
measured from FUSE spectra. Ne = 0.02 was found from an analysis,
including optical spectra \citep{2004A&A...427..685W} and F =
$10^{-5}$  was determined \citep{2005A&A...433..641W}. Preliminary
results for other light element abundances  were presented by
\citet{2007ASPC..372..237R}: Si $< 7.3 \times 10^{-5}$, P = $6.4
\times 10^{-6}$, S = $5.0 \times 10^{-5}$. An upper abundance limit
for argon of roughly solar was derived \citep[Ar $< 8 \times
  10^{-5}$,][]{2007A&A...466..317W}. From a fit to \ion{Fe}{viii}
lines in FUSE spectra, a solar iron abundance was determined
\citep{2011A&A...531A.146W}. 

\begin{table}
\begin{center}
\caption{Final adopted parameters for the program stars \pgev\ and
  \pgfuenf. The results are compared to the prototype of the PG\,1159
  spectral class (\elf) and the Sun.\tablefootmark{a} }
\label{tab:stars} 
\tiny
\begin{tabular}{rrrrr}
\hline 
\hline 
\noalign{\smallskip}
          & \pgev                 &\pgfuenf              & \elf & Sun\tablefootmark{b}\\
\hline 
\noalign{\smallskip}
\Teff/\,K & 150\,000              & 150\,000 &           140\,000 \\
\logg     & 6.5                   & 7.5      &           7.0 \\
\noalign{\smallskip}
H         &  $<0.10$              & $<0.10$              & $<0.02$              & 0.74\\
He        &  0.38                 & 0.43                 & 0.33                 & 0.25\\
C         &  0.57                 & 0.38                 & 0.48                 & $2.4 \times 10^{-3}$\\
N         &  0.015                & $<1.5 \times 10^{-4}$ & 0.001                & $6.9 \times 10^{-4}$\\
O         &  0.016                & 0.17                 & 0.17                 & $5.7 \times 10^{-3}$\\
F         &  $ 1.0 \times 10^{-5}$ & $ 1.0 \times 10^{-4}$ & $ 3.2 \times 10^{-6}$ & $5.0 \times 10^{-7}$\\
Ne        &  0.02                 & 0.02                 & 0.02                 & 0.0013 \\
Na        &  $<0.01$              & ---                  & ---                  & $2.9 \times 10^{-5}$\\
Al        &  $<1.0 \times 10^{-3}$ & $<3.2 \times 10^{-4}$ & $<3.2 \times 10^{-4}$ & $5.6 \times 10^{-5}$\\
Si        &  $ 6.6 \times 10^{-4}$ & $ 6.6 \times 10^{-4}$ & $ 3.6 \times 10^{-4}$ & $6.6 \times 10^{-4}$\\
P         &  $<3.0 \times 10^{-5}$ & $<3.0 \times 10^{-5}$ & $<6.4 \times 10^{-6}$ & $5.8 \times 10^{-6}$\\
S         &  $ 1.0 \times 10^{-4}$ & $ 3.1 \times 10^{-4}$ & $ 3.1 \times 10^{-4}$ & $3.1 \times 10^{-4}$\\
Cl        &  $<1.0 \times 10^{-3}$ & $<1.0 \times 10^{-3}$ & $<1.0 \times 10^{-3}$ & $8.2 \times 10^{-6}$\\
Ar        &  $<8.0 \times 10^{-5}$ & $<8.0 \times 10^{-5}$ & $<3.2 \times 10^{-5}$ & $7.3 \times 10^{-5}$\\
Fe        &  $ 1.3 \times 10^{-3}$ & $ 1.3 \times 10^{-3}$ & $ 1.3 \times 10^{-3}$ & $1.3 \times 10^{-3}$\\
\noalign{\smallskip} \hline
\end{tabular} 
\tablefoot{  \tablefoottext{a}{Abundances in mass fractions and
    surface gravity $g$ in cm\,s$^{-2}$. } \tablefoottext{b}{Solar
    abundances from \citet{2009ARA&A..47..481A}.} Parameters for
  \elf\ from \citet{2007A&A...462..281J} and references therein,
  except for the abundances of Fe \citep[from][]{2011A&A...531A.146W} and
  Al, S, Cl (this work). Upper limits for H in \pgev, \pgfuenf,
  and \elf\ from 
  \citet{1991A&A...247..476W,1991A&A...244..437W,1996A&A...307..860W}.
} 
\end{center}
\end{table}

\section{Observations and line identifications}
\label{sect:observations}

The raw FUSE data for the program stars were retrieved from the
Mikulski Archive for Space Telescopes (MAST). The observation IDs were
P1320201 for PG 1144+005, and P1320101, F0200101, F0200102 for PG
1520+525. Both were observed through the LWRS spectrograph aperture,
and the data were obtained in time-tag mode. We describe the data
  reduction in some detail because the previous reductions in our
  earlier work did not go past the point of coaligning spectra from
  exposures for the eight individual channels separately, whereas now
  the spectra from the individual channels were combined into a single
  spectrum.

The exposures obtained in the observations P1320101 and P1320201 were
photometric in all channels, with exposure-to-exposure flux variations
under 1\%, indicating good alignment, and consistent to within a few
percent when comparing one channel to another. Only one exposure was
successful in each of F0200101 and F0200102, and pointing was unstable
for portions of these exposures; the resulting spectra were normalized
to match the earlier P1320101 data.

Raw data were processed twice with CalFUSE v3.2.3: once with screening
parameters set to extract data only during orbital night, and once to
extract data during orbital day. Zero-point offsets in the wavelength
scale were adjusted for each exposure by shifting each spectrum to
coalign narrow interstellar absorption features. The individual
exposures from each observation were then combined to form composite
day and night spectra for each channel. The day and night spectra were
then compared at the locations of all the known airglow emission
lines. If the day spectra showed any excess flux in comparison to the
night spectra at those wavelengths, the corresponding pixels in the
day spectra were flagged as bad and were not included in subsequent
processing. Significant airglow was only present during orbital night 
at Lyman~$\beta$, and faint emission can be discerned at
Lyman~$\gamma$. This residual emission would affect fits to the
interstellar absorption at these wavelengths, but has no impact on any
of the analyses we present.

Some additional customized processing was performed for \pgfuenf. The
first $\approx 350$ seconds of the first and third exposures were lost
to the limb-angle screening with the default processing
settings. However, an examination of the data showed that the
violation of the limb angle constraint was slight, so we modified the
constraint check to regain over 700 seconds of exposure
time. Similarly, 315 seconds were lost to burst screening in detector
2A during F0200101, and losses in F0200102 ranged from 135 seconds in
detector 2B to 432 seconds in detector 1A. An examination of the raw
data showed no obvious background enhancement from the event burst, so
the processing was performed with a higher burst-detection threshold
to regain the lost exposure time. The net exposure time was 6\,830
seconds for \pgev, with just slight variations from channel to
channel, and ranged from a low of 7\,758 seconds in the SiC1b spectrum
to 8\,600 seconds in the LiF1a spectrum for \pgfuenf.

The final step was to combine the spectra from the four instrument
channels into a single composite spectrum. Because of residual
distortions in the wavelength scale in each channel, additional shifts
of localized regions of each spectrum were required to coalign the
spectra; these shifts were typically only 1--2 pixels. Bad pixels
resulting from detector defects were flagged at this point and
excluded from further processing. Finally, the spectra were resampled
onto a common wavelength scale and combined, weighting by
signal-to-noise ratio (S/N) on a pixel-by-pixel basis.

The final spectrum has S/N = 13--23 per 0.013\,\AA\ pixel for
\pgfuenf, and 14--26 per pixel for \pgev. The effects of fixed-pattern
noise are minimized by the fact that the positions of the spectra on
the detectors varied during each observation, and by the fact that
nearly every wavelength bin was sampled by at least two different
detectors.

The FUSE spectra of both program stars (Figs.\,\ref{fig:pg1520_fuse}
and \ref{fig:pg1144_fuse}) are quite similar to that of the prototype
\elf, as can be expected from the similar atmospheric parameters. They
are dominated by broad lines from \ion{He}{ii}, \ion{C}{iv}, and
\ion{O}{vi, } and exhibit some lines of light metals that will be
discussed below \citep[see][for detailed line lists established for
  the prototype]{2007A&A...462..281J}. In \pgfuenf, we also identify
lines of \ion{O}{v}, which are seen in \elf, too, but which were not
identified in previous work \citep{2007A&A...462..281J}. Detailed
information about \ion{O}{v} lines was compiled for two cool PG\,1159
stars \citep{2015A&A...582A..94W}. \ion{O}{v} is not detected in
\pgev. Ionisation is shifted to higher stages because the surface
gravity is one dex lower than that of \pgfuenf. We also see iron lines
in \elf\ and our program stars \citep[\ion{Fe}{vii} and
  \ion{Fe}{viii},][]{2011A&A...531A.146W}. Other species are discussed
in detail below. The FUSE spectra shown in this paper were wavelength
shifted such that the photospheric lines are at rest wavelengths
according to the measured radial velocities of $15\pm5$~km\,s$^{-1}$
and $10\pm5$~km\,s$^{-1}$ for \pgev\ and \pgfuenf. 

\begin{table}
\begin{center}
\caption{Number of non-LTE levels and lines of model ions used for
  line formation calculations of metals.\tablefootmark{a} }
\label{tab:modelatoms} 
\tiny
\begin{tabular}{ccccccccc}
\hline 
\hline 
\noalign{\smallskip}
   & III  &  IV     &    V     &   VI    &  VII     & VIII  \\
\hline 
\noalign{\smallskip}
C  & 6, 4 & 54, 279  \\   
N  &      & 6, 4    & 54, 297  \\
O  &      & 83, 637 & 105, 671 & 54, 280 \\
F  &      & 1, 0    & 16, 32   & 18, 36  & 14, 28   \\
Ne &      &         &          & 92, 687 & 103, 761 & 77, 510  \\
Na &      & 49, 39  & 52, 229  & 65, 301 & 29, 80   \\ 
Mg &      & 31,93   & 52, 175  & 26, 55  & 46, 147  \\
Al & 1, 0 & 15, 20  & 30, 87   & 43, 126 & 43, 160   \\
Si &      & 30, 102 & 25, 59   & 45, 193 & 61, 138  \\
P  &      & 21, 9   & 18, 12 \\
S  &      &         & 39, 107  & 25, 48  & 38, 120   \\
Cl &34,107& 20, 34  & 13, 10   & 36, 94  & 25, 64    \\
Ar &      &         &          & 8, 8    & 40, 130  & 41, 199  \\
\noalign{\smallskip} \hline
\end{tabular} 
\tablefoot{ \tablefoottext{a}{First and second number of each table
    entry denote the number of levels and lines, respectively. Not
    listed for each element is the highest considered ionisation stage, which
    comprises its ground state only. See text for the treatment of
    iron.}  } 
\end{center}
\end{table}

\begin{figure*}[bth]
 \centering  \includegraphics[width=0.95\textwidth]{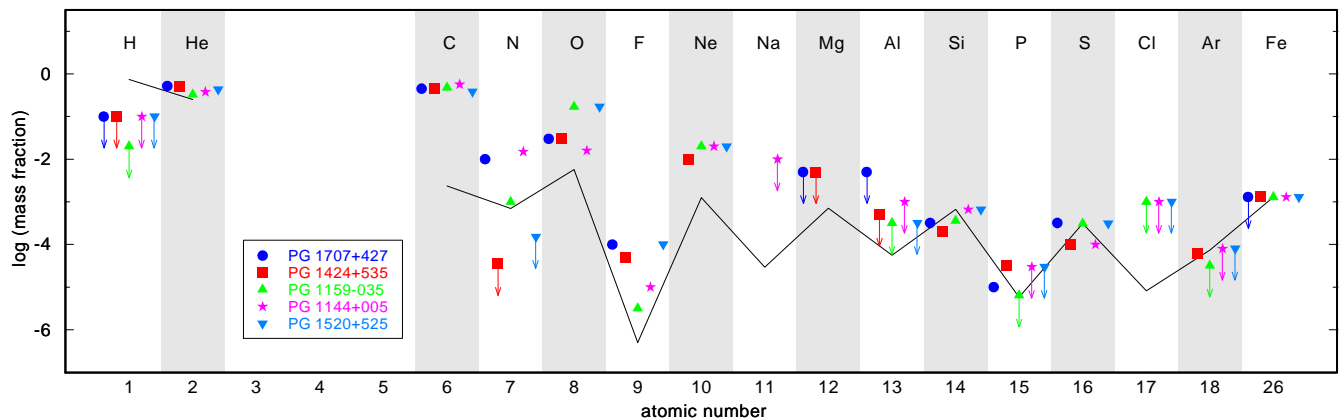}
  \caption{Element abundances measured in five PG\,1159 stars. Upper
    limits are indicated by arrows. Data for our two program stars and
    \elf\ are from Tab.\,\ref{tab:stars}. Data for \pgsieben\ and
    \pgvier\ are from \citet{2015A&A...582A..94W} except for the H
    abundances \citep[from][]{1991A&A...244..437W}. The black solid
    line indicates solar abundances.}\label{fig:abu}
\end{figure*}

To identify lines from the interstellar medium (ISM) and to judge
their potential contamination of photospheric lines, we used the
program OWENS \citep{2002P&SS...50.1169H,2003ApJ...599..297H}.
This can consider different clouds with individual radial and turbulent
velocities, temperatures, column densities, and chemical compositions.
In the FUSE observations of both stars, we identified ISM lines of
\ion{H}{i},
\ion{D}{i},
H$_2$,
\ion{C}{i--iii},
\ion{N}{i--ii},
\ion{O}{i},
\ion{Si}{ii},
\ion{P}{ii},
\ion{S}{iii},
\ion{Ar}{i}, and
\ion{Fe}{ii}.
The blue graphs in 
Figs.\,\ref{fig:pg1520_fuse} and
       \ref{fig:pg1144_fuse} show the photospheric models including the ISM lines.

Interstellar reddening is $E(B-V) = 0.015\pm0.010$ for both stars,
determined by comparing model fluxes normalized to measured 2MASS
infrared colours \citep[$K_s = 16.282$ for \pgev\ and $J = 16.381$ for
  \pgfuenf,][]{2011ApJS..197...38D} with the continuum flux and
spectral shape of the FUSE spectra.

\section{Model atoms and model atmospheres}
\label{sect:models}

We used the T\"ubingen Model Atmosphere Package (TMAP) to compute
non-LTE, plane-parallel, line-blanketed atmosphere models in radiative
and hydrostatic equilibrium
\citep{1999JCoAM.109...65W,2003ASPC..288...31W}. They include the
three most abundant elements, namely He, C, and O. All other species
were treated, one by one, as trace elements, i.e., keeping  the
atmospheric structure fixed. In the same manner, an extended model atom for
C was introduced, meaning that non-LTE population numbers were
computed for highly-excited levels, which were treated in LTE during
the preceding model-atmosphere
computations. Table\,\ref{tab:modelatoms} summarizes the number of
considered non-LTE levels and radiative transitions between them. All
model atoms were built from the publicly available T\"ubingen Model
Atom Database
(TMAD\footnote{\url{http://astro.uni-tuebingen.de/~TMAD}}, constructed
as part of the German Astrophysical Virtual Observatory, GAVO),
which is comprised of data from different sources, namely
\citet{1975aelg.book.....B}, the databases of the National Institute
of Standards and Technology
(NIST\footnote{\url{http://www.nist.gov/pml/data/asd.cfm}}), the
Opacity
Project\footnote{\url{http://cdsweb.u-strasbg.fr/topbase/topbase.html}}
\citep[OP,][]{1994MNRAS.266..805S},
CHIANTI\footnote{\url{http://www.chiantidatabase.org}}
\citep{1997A&AS..125..149D,2013ApJ...763...86L}, as well as the
Kentucky Atomic Line
List\footnote{\url{http://www.pa.uky.edu/~peter/atomic}}. 
We put
significant effort into the development and testing of model atoms for
species that had  not been investigated before in hot PG1159 stars (Na, Mg, Al, Cl)
with an emphasis on the highest ionization stages.

For iron we used a statistical approach, employing typically seven
superlevels per ion linked by superlines, together with an opacity
sampling method \citep{1989ApJ...339..558A,2003ASPC..288..103R}.
Ionization stages \ion{Fe}{vi--ix} augmented by a single, ground-level
stage \ion{Fe}{x} were considered. We used the complete line
list of Kurucz \citep[so-called LIN lists, comprising about
  $1.2\times10^6$ lines of the considered
  ions;][]{kurucz1991,kurucz2009, kurucz2011} for the computation of
the non-LTE population numbers, and the so-called POS lists (that
include only the subset of lines with well known, experimentally
observed line positions) for the final spectrum synthesis.

\section{Results}
\label{sect:results}

For the abundance analysis, we adopted the effective temperature and
surface gravity values of both stars that had been determined in previous
work. We found no hint from the FUSE spectra that these values need
improvement. This is corroborated by the relative strengths of
\ion{Si}{v} and \ion{Si}{vi} lines which are well matched  by the
models. The new detection of many \ion{O}{v} lines in
\pgfuenf\ together with \ion{O}{vi} lines additionally confirms that
the basic photospheric parameters, \Teff\ and \logg, are well
established. The same holds for the abundance of the main atmospheric
constituents (He, C, and O).

In the following, we discuss the abundance determinations for trace
elements in detail. Results for the two program stars are given in
Table\,\ref{tab:stars} and compared to the prototype \elf. They are
also displayed graphically in Fig.\,\ref{fig:abu} in comparison to
\elf\ and two other previously analysed PG\,1159 stars
\citep[\pgsieben\ and \pgvier,][]{2015A&A...582A..94W}. Typical errors
are 0.3 dex.

\paragraph{Nitrogen.}

The relatively high N abundance in \pgev\ was found from
optical spectroscopy and cannot be further assessed with the FUSE
spectrum. The strongest lines in the model are \ion{N}{v} 4d--6f and
4f--6g at 1048.2\,\AA\ and 1049.7\,\AA, but they are blended by strong
interstellar lines in the observation. The next strongest feature is
the 4p--6d doublet at 1019.0/1019.3\,\AA, but it is blended by a
strong photospheric \ion{O}{vi} line and an interstellar line. For the
same reason, the upper N limit for \pgfuenf,\ derived previously from
the \ion{N}{v} resonance doublet, cannot be improved.

\paragraph{Fluorine, neon, argon, iron.}

The previously determined abundances of F, Ne, Ar (upper limits), and
Fe in both stars are confirmed.

A weak \ion{Ne}{vi} line appearing in the models at 1186.6\,\AA\ is
possibly present in \pgev. The line is not listed in the NIST and
Kentucky databases but in the CHIANTI database. The oscillator
strength given there is $f=0.068$, while we use the larger OP value of
$f=0.097$ in our models. A stronger \ion{Ne}{vi} line in the models,
that is also listed in the NIST database, is seen at
972.93\,\AA\ (near the \ion{Ne}{vii} 973.33\,\AA\ line). However, it
is blended by strong interstellar lines.

The model for \pgev\ displays the 5f--6g and 5g--6h lines of \ion
{Ar}{viii} at 1155.0 and 1164.1\,\AA\ \citep{1997JQSRT..58..627H}.
However, their wavelength positions are not  well enough known 
($\pm 0.84$ and $\pm 1.2$\,\AA, Kentucky database) to identify them
with features in the observed spectra. We speculated about the
presence of these argon lines in the hottest PG\,1159 stars
\citep[e.g., K1--16,][]{2007A&A...474..591W}.

\paragraph{Sodium.}

Our models predict detectable, i.e., strong enough lines of
\ion{Na}{v} and \ion{Na}{vi} only at high abundances of the order
0.01. Even worse, the line positions are not  sufficiently accurate. In the model for \pgev, a weak \ion{Na}{vii} line with well
known wavelength position (1114.08\,\AA, NIST) appears, but no tight
abundance limit can be derived (Na $<$ 0.01). Neither the model for
\pgfuenf\  exhibits this line because of its higher gravity, nor
the model for \elf,\ because of its lower temperature.

\begin{figure}[t]
 \centering  \includegraphics[width=1.0\columnwidth]{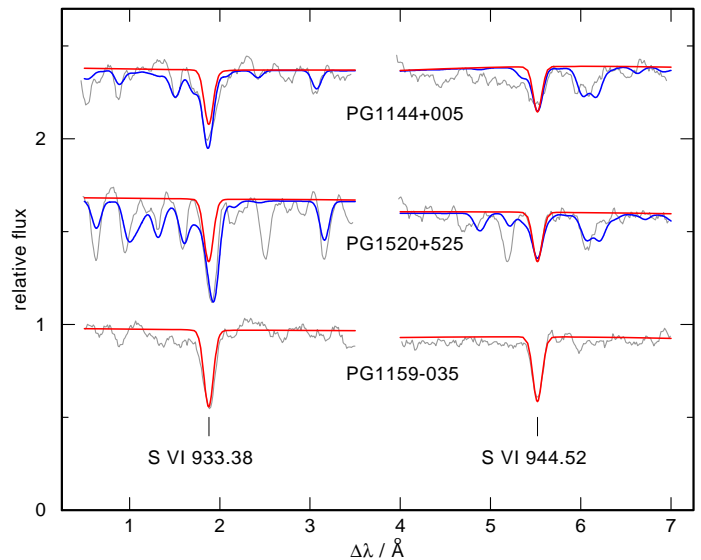}
  \caption{Model fits to the \ion{S}{vi} resonance doublet in the two
    program stars and in the prototype \elf, with parameters as given
    in Table\,\ref{tab:stars}. In \pgev\ and \pgfuenf, the
    933.38\,\AA\ component is blended by an interstellar H$_2$ line. Red
    curves: pure photospheric spectrum. Blue curves: ISM lines
    included in the model (not for \elf).}\label{fig:s6}
\end{figure}

\paragraph{Magnesium.}

Our models predict \ion{Mg}{v} lines which should be detectable if the
abundance is higher than about $10^{-3}$. Unfortunately, these lines
are not listed in NIST and Kentucky databases so that their wavelength
position seems  uncertain. Consequently, not even an upper
limit can be derived. The same holds for a single \ion{Mg}{vii} line
in the model for \pgev. 

\paragraph{Aluminum.}

Our models predict weak lines from \ion{Al}{v} and \ion{Al}{vi} that
can be used to assess the Al abundance. The strongest computed lines
of these ions are \ion{Al}{v} 1068.26\,\AA\ and 1088.67\,\AA, that are
the strongest components of $^4$P$^{\rm o}$--$^4$D and $^4$D$^{\rm
  o}$--$^4$F multiplets, and \ion{Al}{vi} 1056.00\,\AA, the strongest
component of a $^5$P--$^5$D$^{\rm o}$ multiplet. In \pgfuenf,
weak line features are at these wavelength positions. However, we regard
these identifications as uncertain. An upper limit of Al $<3.2\times
10^{-4}$ (about six times solar) can be derived. In the model for
\pgev, the ionisation of Al shifts to higher stages because of the
lower gravity (hence lower particle densities). Therefore, the derived
upper limit is less strict, namely Al $<1.0\times 10^{-3}$. The Al
abundance in the prototype, \elf, has hitherto not been investigated. Using
an appropriate model atmosphere, we derive Al $<3.2\times 10^{-4}$.

\paragraph{Silicon.}

We identified a few weak lines from \ion{Si}{v} (975.8, 1118.8\,\AA)
and \ion{Si}{vi} (1131.0, 1142.4, 1148.6, 1152.8\,\AA). We obtain best-model fits at a
solar abundance value (Si = $6.6 \times 10^{-4}$) for both stars,
hence, we do not confirm the stricter upper limits derived by
\citet{2007ASPC..372..237R} from the absence of \ion{Si}{iv} lines in
the FUSE observations.

\paragraph{Phosphorus.}

From the nondetection of the \ion{P}{v} resonance doublet (1118,
1128\,\AA), we derive an upper limit of P $< 3.0 \times 10^{-5}$ for
both stars. This value is more conservative than the limits claimed by
\citet{2007ASPC..372..237R}.

\paragraph{Sulphur.}

From the \ion{S}{vi} 933/944\,\AA\ resonance doublet
(Fig.\,\ref{fig:s6}) and from two much weaker subordinate lines at
1000.5\,\AA\ and 1117.8\,\AA, we find S = $1.0 \times 10^{-4}$ and $3.1
\times 10^{-4}$ for \pgev\ and \pgfuenf, respectively, being solar
within error limits. This result is in stark contrast to the earlier
results reported by \citet{2007ASPC..372..237R} that indicate a one
dex subsolar abundance for these stars. We discovered that the low S
values resulted from an error in the \ion{S}{vi} model atom (wrong
statistical weights for two energy levels) employed for the
computation of the non-LTE population numbers. 

We mention that the error in the model atom also affected the sulphur
abundance determination in the prototype \elf\ in our earlier work
\citep{2007A&A...462..281J}. A surprisingly low value of just 2\%
solar was found, whereas a reanalysis with our corrected model
atom arrives at a solar abundance.

\paragraph{Chlorine.}

Lines of \ion{Cl}{vii} enable us to put an upper abundance limit, though
it is not very tight. \ion{Cl}{vii} is a one-valence-electron system. The
strongest predicted lines are from the 4d--5p transition
(997.03\,\AA\ and 999.32\,\AA) and the 5f--7g transition at
944.1\,\AA\  but, for the latter, the wavelength uncertainty is too large
(0.57\,\AA; Kentucky database). From the former we find Cl $<10^{-3}$
for both program stars as well as for \elf.

\section{Summary and discussion}
\label{sect:discussion}

We have analysed FUV spectra of two hot PG\,1159 stars (\pgev\ and
\pgfuenf, both \Teff = 150\,000~K) with the focus on element
abundance determinations. Previously obtained values for \Teff\ and
\logg\ were confirmed. With our improved model atmospheres, we additionally
derived new findings on the prototype itself (\elf; 140\,000~K). The
results for these three objects are summarized in
Table~\ref{tab:stars}. In Fig.\,\ref{fig:abu}, the element
abundances are displayed, together with the results for two cooler
PG\,1159 stars \citep[\pgsieben\ and \pgvier\ with 85\,000 and
  110\,000~K;][]{2014A&A...569A..99W}. Taken together, these five
objects represent the only PG\,1159 stars with comprehensive element
abundance determinations.

\begin{table}
\begin{center}
\caption{Asteroseismic, spectroscopic, and initial masses of the five considered PG\,1159 stars. }
\label{tab:masses} 
\begin{tabular}{llll}
\hline 
\hline 
\noalign{\smallskip}
Star   & M$_{\rm puls}$/M$_\odot$ & M$_{\rm spec}$/M$_\odot$& M$_{\rm init}$/M$_\odot$\tablefootmark{(e)} \\
\hline 
\noalign{\smallskip}

\pgvier   & --                                      & $0.51^{+0.07}_{-0.01}$\tablefootmark{(c)} & $\approx$1--2 \\
\noalign{\smallskip}
\pgsieben & $0.542^{+0.014}_{-0.012}$\tablefootmark{(a)} & $0.53^{+0.17}_{-0.03}$\tablefootmark{(c)} & $\approx$1--3 \\
\noalign{\smallskip}
\elf      & $0.565^{+0.025}_{-0.009}$\tablefootmark{(b)} & $0.54^{+0.05}_{-0.01}$\tablefootmark{(c)} & $\approx$1--2 \\
\noalign{\smallskip}
\pgev     & --                                      & $0.56^{+0.07}_{-0.03}$\tablefootmark{(d)} & $\approx$1--2.5 \\
\noalign{\smallskip}
\pgfuenf  & --                                      & $0.62^{+0.15}_{-0.08}$\tablefootmark{(d)} & $\approx$1--4 \\
\noalign{\smallskip} \hline
\end{tabular} 
\tablefoot{ 
\tablefoottext{a}{\citet{2008A&A...478..869C};}
\tablefoottext{b}{\citet{2009A&A...499..257C};}
\tablefoottext{c}{\citet{2014A&A...569A..99W};}
\tablefoottext{d}{this work;}
\tablefoottext{e}{estimated from initial-to-final mass relation of \citet{2000A&A...363..647W} using M$_{\rm spec}$.}
 } 
\end{center}
\end{table}

For the comparison of the results with evolutionary models, stellar
masses are relevant. In Table~\ref{tab:masses} we list the
spectroscopic masses derived by comparing them with the observed values of
\Teff\ and \logg\ with evolutionary tracks by
\citet{2009ApJ...704.1605A}. For the two pulsators in the sample,
asteroseismic masses are also available, being in good agreement with
the spectroscopic values. Also listed are the masses of the
main-sequence progenitors, as estimated from the initial-to-final mass
relation of \citet{2000A&A...363..647W}, taking into account the
uncertainties of the spectroscopic masses.

The previously determined abundances of the main atmospheric
constituents (He, C, O, Ne) in our two program stars were
confirmed. The same holds for N, F, Ar, and Fe. The investigation on
Si, P, and S gave improved results compared to earlier, preliminary
work. For the first time, we have  assessed the abundances of Na,
Al, and Cl, and derived upper limits. 

We used our model atmosphere grid to reassess the abundances of trace
elements in the prototype \elf. The main result is an improved S
abundance, which is solar in contrast to the strong S depletion (0.02
solar) claimed by \citet{2007A&A...462..281J}, which was caused by an
error in the model atom. We also derived upper limits for Na, Al,
and Cl for this star.

The results can be compared to evolutionary models by
\citet{2013MNRAS.431.2861S}, who presented intershell abundances for
stars with initial masses of 1.8--6~$M_\odot$. For \elf, the authors
found good agreement for the species investigated (He, C, O, F, Ne,
Si, P, S, Fe) with the exception of sulphur. As pointed out, this
discrepancy is now resolved. For the two cool PG\,1159 stars mentioned
above, good agreement of observed and predicted element abundances was
found \citep{2014A&A...569A..99W}. As to our two program stars, all
element abundances are similar to those found for the other three
PG\,1159 stars (Fig.\,\ref{fig:abu}) and, hence, confirm the
evolutionary models. For a thorough discussion, we can therefore refer
to \citet{2014A&A...569A..99W}, however, N, Na, Al, and Cl deserve
particular attention here. The Na and Cl abundances in PG\,1159
stars were examined for the first time.

\pgev\ is remarkable because of its relatively large N abundance. This
phenomenon is shared by some other PG\,1159 stars (like the mentioned
cooler object \pgsieben) and is attributed to the fact that these
objects  suffered their last thermal pulse when they already were
white dwarfs (so-called very late thermal pulse, VLTP).

Only for \pgev\ could a useful upper limit for the sodium abundance  be
derived, which is 341 times oversolar. Karakas \& Shingles
(priv. comm.) report that the Na intershell abundance at the last
thermal pulse in the 1.8 and 3~$M_\odot$ models presented in
\citet{2013MNRAS.431.2861S} has increased considerably to 23.1 and
18.1 times oversolar after the last thermal pulse. This is still well
below our detection limit.

For aluminum, upper limits of 18 and 6 times solar were derived for
\pgev\ and \pgfuenf. The Al intershell abundance at the last thermal
pulse in the mentioned 1.8 and 3~$M_\odot$ models is 1.5 and 2 times
the initial solar value. Hence, the rather modest production of Al is
confirmed by our result.

For the two program stars, as well as for \elf, only a high upper
limit of Cl $= 10^{-3}$ was derived, which is about two dex
oversolar. The intershell abundance in the 1.8~$M_\odot$ and
3~$M_\odot$ models has increased to 2.3 and 2.1 times oversolar after
the last thermal pulse, i.e., well below our detection limit.

To summarize, the abundances of elements up to iron in PG\,1159 stars
is in good agreement with intershell abundances predicted by
stellar-evolution models. For a number of interesting species (Na,
Al, Cl), however, only upper limits can be derived by spectral
analyses. It turned out that for the relevant ionisation stages ({\sc
  iv--vii}), precise atomic data is lacking, impeding line
identifications. This obstacle can only be overcome by laboratory
measurements. The detailed investigation into the trace element abundances
as a function of stellar mass is additionally hindered by the
relatively large uncertainty in the spectroscopic mass determination,
which itself is due to the uncertainty of the surface gravity
measurement ($\pm 0.5$~dex). Considerable improvement is expected from
accurate parallaxes that will be provided by GAIA.

\begin{acknowledgements} 
We thank Amanda Karakas and Luke Shingles for reporting unpublished
results to us.  T. Rauch is supported by the German Aerospace Center
(DLR) under grant 50\,OR\,1507. The TMAD service
(\url{http://astro-uni-tuebingen.de/~TMAD}) used to compile atomic
data for this paper was constructed as part of the activities of the
German Astrophysical Virtual Observatory. This research has made use
of the SIMBAD database, operated at CDS, Strasbourg, France, and of
NASA's Astrophysics Data System Bibliographic Services. Some of the
data presented in this paper were obtained from the Mikulski Archive
for Space Telescopes (MAST). This work had been done using the profile
fitting procedure OWENS, developed by M\@. Lemoine and the FUSE French
Team. 
\end{acknowledgements}

\bibliographystyle{aa}
\bibliography{aa}

\end{document}